\documentclass[pre,a4paper,onecolumn,notitlepage,nofootinbib]{revtex4-1}
\usepackage{amsmath}
\usepackage{bbm}
\usepackage{bm}
\usepackage{latexsym}
\usepackage{amsfonts}

\usepackage[shortlabels]{enumitem}
\usepackage{multirow}
\usepackage[table]{xcolor}

\usepackage[pdftex]{graphicx}
\usepackage{subfigure}
\usepackage{epstopdf}
\usepackage{color} 
\usepackage{booktabs} 
\usepackage[utf8]{inputenc}
\usepackage[T1]{fontenc}
\usepackage{times}

\newcommand{\beq}{\begin{equation}}
\newcommand{\eeq}{\end{equation}}

\begin{document}

\title{\bf Classifying attention deficit hyperactivity disorder in children with non-linearities in actigraphy}

\author{Jeremi K. Ochab}
\email{jeremi.ochab@uj.edu.pl}
\homepage[Visit: ]{http://cs.if.uj.edu.pl/jeremi/index_EN.html}
\address{Marian Smoluchowski Institute of Physics and Mark Kac Center for Complex Systems Research, Jagiellonian University, ul. {\L}ojasiewicza 11, 30---348 Krak\'ow, Poland}
\author{Krzysztof Gerc}\email{krzysztof.gerc@uj.edu.pl}
\address{Developmental and Health Psychology Department, Institute of Applied Psychology, Faculty of Management and Social Communication, Jagiellonian University, ul. {\L}ojasiewicza 4, 30---348 Krak\'ow, Poland}
\author{Magdalena F{a}frowicz}
\address{Department of Neurobiology, Neuroimaging Group, Malopolska Center of Biotechnology, Jagiellonian University, ul. Gronostajowa  7a, 30---348 Krak\'ow, Poland}
\author{Ewa Gudowska-Nowak}
\address{Marian Smoluchowski Institute of Physics and Mark Kac Center for Complex Systems Research, Jagiellonian University, ul. {\L}ojasiewicza 11, 30--348 Krak\'ow, Poland}
\address{Malopolska Center of Biotechnology, Jagiellonian University, ul. Gronostajowa 7, 30---348 Krak\'ow, Poland}
\author{Tadeusz Marek}
\address{Department of Cognitive Neuroscience and Neuroergonomics, Institute of Applied Psychology, Jagiellonian University, ul. {\L}ojasiewicza 4, 30---348 Krak\'ow, Poland}
\author{Maciej A. Nowak}
\address{Marian Smoluchowski Institute of Physics and Mark Kac Center for Complex Systems Research, Jagiellonian University, ul. {\L}ojasiewicza 11, 30---348 Krak\'ow, Poland}
\author{Halszka Ogi\'nska}
\address{Department of Cognitive Neuroscience and Neuroergonomics, Institute of Applied Psychology, Jagiellonian University, ul. {\L}ojasiewicza 4, 30---348 Krak\'ow, Poland}
\author{Katarzyna Ole\'s}
\address{Marian Smoluchowski Institute of Physics, Jagiellonian University, ul. {\L}ojasiewicza 11, 30---348 Krak\'ow, Poland}
\author{Ma{\l}gorzata Rams}
\address{Faculty of Physics, Astronomy and Applied Computer Science, Jagiellonian University, ul. {\L}ojasiewicza 11, 30-348 Krak\'ow, Poland}
\author{Jerzy Szwed}
\address{Marian Smoluchowski Institute of Physics and Mark Kac Center for Complex Systems Research, Jagiellonian University, ul. {\L}ojasiewicza 11, 30---348 Krak\'ow, Poland}
\author{Dante R. Chialvo}
\address{Center for Complex Systems \& Brain Sciences (CEMSC3), Universidad Nacional de San Mart\'{i}n,25 de Mayo 1169, San Mart\'{i}n, (1650), Buenos Aires, Argentina}\affiliation{Consejo Nacional de Investigaciones Cient\'{i}ficas y Tecnol\'{o}gicas (CONICET), Godoy Cruz 2290, Buenos Aires, Argentina}


\date{\today}

\begin{abstract}
\begin{description}
\item[Objective]
This study provides an objective measure based on actigraphy for Attention Deficit Hyperactivity Disorder (ADHD) diagnosis in children. We search for motor activity features that could allow further investigation into their association with other neurophysiological disordered traits.
\item[Methods]
The study involved $n=29$ (48 eligible) male participants aged $9.89\pm0.92$ years (8 controls, and 7 in each group: ADHD combined subtype, ADHD hyperactive-impulsive subtype, and autism spectrum disorder, ASD) wearing a wristwatch actigraph continuously for a week ($9\%$ losses in daily records) in two acquisition modes. We analyzed 47 quantities: from sleep duration or movement intensity to theory-driven scaling exponents or non-linear prediction errors of both diurnal and nocturnal activity. We used them in supervised classification to obtain cross-validated diagnostic performance.
\item[Results]
We report the best performing measures, including a nearest neighbors 4-feature classifier providing $69.4\pm1.6\%$ accuracy, $78.0\pm2.2\%$ sensitivity and $60.8\pm2.6\%$ specificity in a binary ADHD vs control classification and $46.5\pm1.1\%$ accuracy (against $25\%$ baseline), $61.8\pm1.4\%$ sensitivity and $79.30 \pm0.43\%$ specificity in 4-class task (two ADHD subtypes, ASD, and control). The most informative feature is skewness of the shape of Zero Crossing Mode (ZCM) activity. Mean and standard deviation of nocturnal activity are among the least informative.
\item[Conclusions]
Actigraphy causes only minor discomfort to the subjects and is inexpensive. The range of existing mathematical and machine learning tools also allow it to be a useful add-on test for ADHD or differential diagnosis between ADHD subtypes. The study was limited to a small, male sample without the inattentive ADHD subtype.
\end{description}
 
\medskip

\noindent {\em Keywords:\/} ADHD, Attention Deficit Hyperactivity Disorder, Supervised Machine Learning, non-linear prediction, Differential Diagnosis, Computer-Assisted Diagnosis, Decision Support Techniques;

\end{abstract}

\maketitle

\section{Introduction}
Mathematical analysis of physiological signals, human movement variability among them, for decades attracted considerable attention of the scientific community, resulting in a large body of available methods (see, references \cite{Akay2000,Stergiou2016book}, which are still not exhaustive). This paper takes advantage of this aggregated wealth, gathers multiple quantifiable characteristics of human actigraphy records, and takes them as input for supervised machine-learning algorithms in order to help in diagnosing and differentiating Attention Deficit Hyperactivity Disorder (ADHD) in children.

Actigraphs are wristwatch-like devices that use accelerometers to register locomotor activity. The analysis of the data they provide is frequently considered a cost-effective method of first choice, used both in clinical research and practice. Since motor activity levels and circadian rhythms are related to a range of psychiatric disorders~\cite{Teicher1995}, their monitoring can be utilized in diagnosis, prediction of treatment response or for sleep assessment. Beside sleep and major mood disorders some neurological diseases such as Parkinson’s disease~\cite{Pan2007,Pan2013a,Sun2013}, vascular dementia~\cite{Pan2013b}, Alzheimer’s disease~\cite{Teicher1995,vanSomeren1996}, schizophrenia~\cite{Nakamura2012} or chronic pain~\cite{Long2008} have been shown to be related to abnormal activity symptoms.

In terms of nosological classification,  ADHD is a separate disease entity, whose diagnostic criteria are included in DSM-5~\cite{DSM5} and ICD-10~\cite{ICD10}. The behavioral symptoms included as criteria for ADHD are attention disorders, psychomotor hyperactivity, and impulsiveness. Attention disorders are predominantly connected to a lowered ability to concentrate and to keep attention on a given task, which include, lowered selectivity as well as excessive susceptibility to distractors. Hyperactivity manifests itself by heightened motor activity, usually chaotic, and undertaken without precise aim or purpose. Impulsiveness on the other hand is connected to a person behaving to stimuli occurring at a given time, however, without reflecting on its adequacy or potential effects. Depending on the intensity of the motor and attention focus symptoms, DSM-5 differentiates between three subtypes of ADHD:
\begin{itemize}
\item	ADHD-predominantly inattentive type - presents with being easily distracted, forgetful, daydreaming, disorganization, poor concentration, and difficulty completing tasks;
\item	ADHD-predominantly hyperactive-impulsive type, henceforth denote for short ‘hi-ADHD’ – presents with excessive fidgetiness and restlessness, hyperactivity, difficulty waiting and remaining seated, and immature behavior, where destructive behaviors may also be present;
\item	ADHD combined type, henceforth denoted ‘c-ADHD’ – is a combination of the first two subtypes.
\end{itemize}
In general, the results of neuropsychological tests used in ADHD research rely on various simultaneous cognitive functions, which as a rule, do not indicate the specific neuronal circuits which malfunction in ADHD persons~\cite{sergeant2002specific}. However, such opportunities have been presented by some research paradigms developed by combining experimental psychology and neurobiological approaches together with advanced mathematical methods of bioelectrical brain activity measurements, and other experimental and neuroimaging techniques. These paradigms include among others, auditory and visual oddball tasks~\cite{folstein2008influence,johnstone2007behavioural,senderecka2012event}, as well as actigraphy studies.  A review of previous research on actigraphy records revealed~\cite{melegari2016actigraphic,de2016use,wiebe2013sleep,moreau2014sleep} that ADHD children were shown to exhibit increased motor activity during sleep and increased nocturnal activity in studies using actigraphy. At the same time, it was reported~\cite{poirier2015night} that the results concerning the nature of the sleep and activity problems in ADHD children are unclear due to methodological limitations of the studies. A large part of what is currently known about attention-deficit/hyperactivity disordered children’s functioning in sleep is based on ADHD Combined Type samples, but there is little research focused on experimentally differentiating the other ADHD subtypes~\cite{wiebe2013sleep,becker2016sleep}, where  behavioral data analysis can be verified with relatively sensitive measurements, and interpreted with the use of sophisticated statistical models.

A common approach to analyze such a record is to inspect activity profiles or histograms, extract circadian frequency and phase shifts, sleep duration and wake after sleep onset or to reduce the time series to a summary statistic such as mean activity level, ratio of nocturnal and diurnal activity, or intra- and interdaily variability, etc.~\cite{Teicher1995,vanSomeren1996,Oginska2014,Matuzaki2014}.
While such measures allow employing classic statistical methods, more relevant information may be exposed when other advanced tools of quantification, like fractal analysis, are applied~\cite{Pan2007,Pan2013a,Sun2013,Holloway2014,Gudowska2016jstat,Wohlfahrt2013}. Specifically, the power law exponent describing temporal autocorrelations of activity correlates with the symptoms of some disorders or conditions. Similar scaling laws have been found in activity period distributions in humans suffering from major depressive disorders~\cite{Nakamura2007,Nakamura2008} and individuals subject to sleep deprivation~\cite{Ochab2014,Gudowska2016jstat,Holloway2014}. Disruption of these laws in the context of rest and activity fluctuations in light and dark phases of the circadian cycle was further studied by~\cite{Proekt2012}. 

In this paper we refrain from delving into hypothetical sources and theoretical repercussions of these results. The focus is on which quantities and to what extent can avail diagnosis of hyperactivity disorders in children. In Methods Section we lay out the relevant details on the measurements and the subjects, as well as on some simple stages of preprocessing the collected data. Next, we list and discuss all the quantities applied for characterizing the individual time series. We explain meaning and details of computing these quantities only briefly, whenever we can refer to existing studies. Finally, Results Section reports on the results of classifying the subjects into different groups. These results are followed with a discussion and conclusions in the last section.

\section{Methods}
\subsection{Study design}
Participants were registered and diagnosed in respective clinics before commencement of the study. Participants and their parents were informed about the procedure and goals of the study, and provided their written consent for the actigraphic and supplementary psychological examination (including WISC-R). Participant selection was based on medical and psychological records. Next, a 7 day actigraphic recording was performed. The key quantities (classification features) were designed before obtaining the data; the cross-validation paradigm was designed after the data had been collected based on their size. The study was approved by the Bioethics Commission at Jagiellonian University.
In order to minimize any potential discomfort, anxiety and other psychological costs related to participation in the study, the caregivers could phone and consult a member of the team (clinical psychologist) at every stage of the experiment, they were provided detailed instructions, and encouraged to ask any questions and express any concerns.

\subsection{Participants}

The ADHD participants were identified in registry of a local (Skawina near Kraków, Poland) specialized educational psychology clinic. The ASD participants were identified at a local (Krak\'ow, Poland) autism and developmental disorders clinic and had been diagnosed by a medical doctor (psychiatrist) and a clinical psychologist. In this study, including the ASD group serves mainly as a way of testing robustness of the classifiers’ specificity in a scenario with multiple disorders to diagnose. The control group was randomly selected from among boys attending regular primary schools located in the same area from where the subjects with developmental disorders lived. The participants had not undergone any pharmacotherapy up to six months prior to the study.

The exclusion criteria before actigraphy recording were: ambiguous ADHD symptoms, suspected co-occurring developmental disorders (ADHD and ASD groups), below normal intelligence assessment (all groups). The control group was comprised of children showing no psychological disorders, nor any other serious somatic diseases; they also came from families that did not reveal pathological traits or characteristics.

The ADHD diagnosis was obtained in accordance with the current standards of assessment~\cite{DSM5,ICD10}. It involved an expert assessing the intensity levels of symptoms in the areas of hyperactivity, impulsiveness and attention deficit, according to the diagnostic criteria of DSM-5~\cite{DSM5}. The assessment was based on clinical observations, as well as, information obtained from the parents and the child during a diagnostic interview. Intellectual functioning of all subjects was determined with the WISC-R scale.
The study involves and relates to the children with the ADHD combined type and the hyperactive-impulsive type only. Due to developmental aspects, it would be inappropriate to the inattentive ADHD type, as children in this group aged 8-10,  have a tendency to alter the intensity of the dominant symptoms of ADHD, as reported in~\cite{kaplan1988synopsis,klykylo2006clinical,lewis2002child,semple2013oxford,lask2003practical}.

\subsection{Actigraphy data}
All subjects were given a motor activity recording device – Micro Motionlogger SleepWatch (Ambulatory Monitoring, Inc., Ardsley, NY), to be worn continuously (without being taken off at night or during bathing) on the wrist of the non-dominant hand. Recording time took no longer than 7 days. The subjects did not perform any additional tasks or tests, and participated in their normal daily activities during this period. The parents were instructed to inform the research team of any infections of the participants and any medication taken during the recording. The actigraph activity monitors were personally put on and taken off each subject by a member of the research team. Three hyperactive-impulsive participants reported displeasure due to wearing the devices during sleep.
\begin{figure}[htbp!]
\centering
\includegraphics[width=0.98\textwidth]{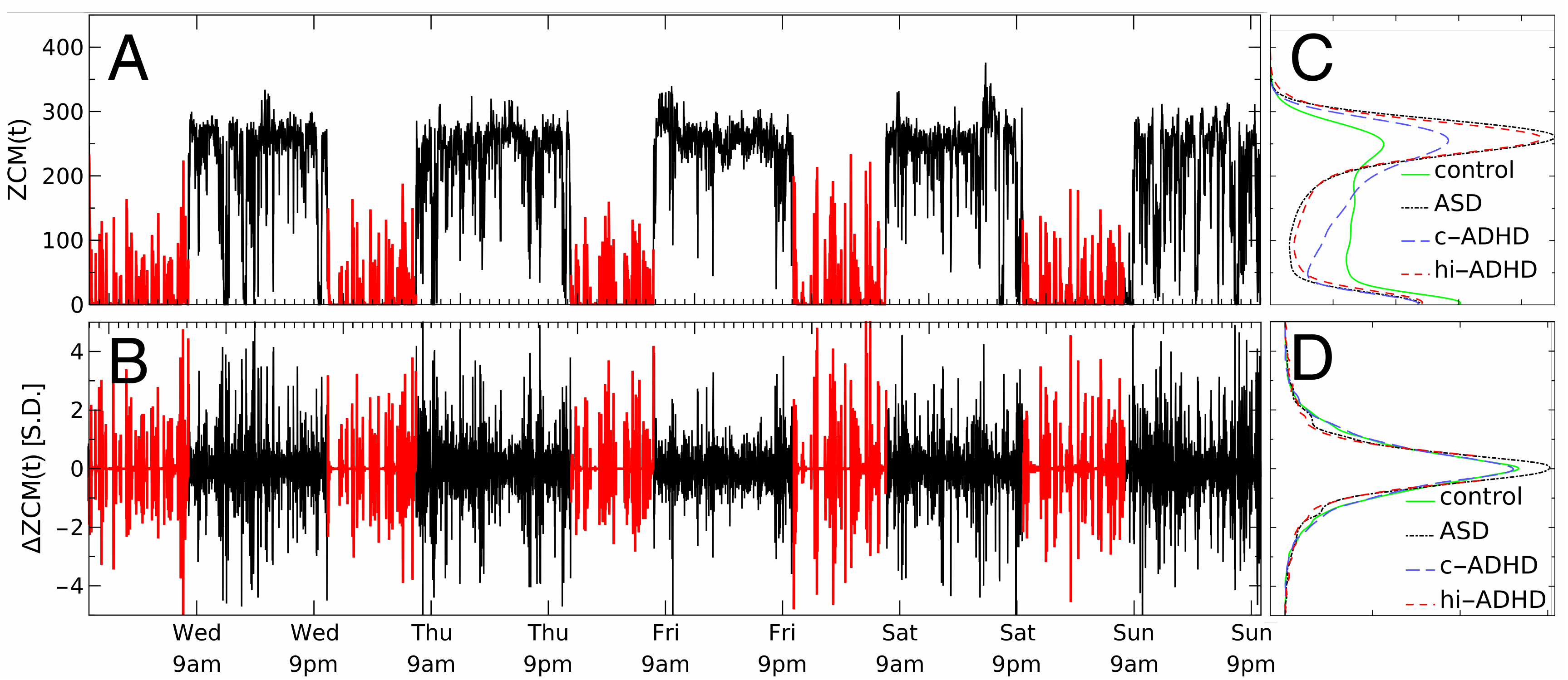}
\caption{\label{fig:raw}
(A) Raw Zero Crossing Mode (ZCM) measurements and (B) standardized increments $\Delta\mathrm{ZCM}$ for an hi-ADHD subject, with black and red indicating diurnal and nocturnal activity. Right panels (C, D) show histograms of activity for the most distinctive subjects in all four groups; horizontal axes indicates probability density, vertical axes are the same as in panels (A, B).}
\end{figure}
The data were collected continuously in 1-minute epochs with the Zero Crossing Mode (ZCM), see Fig.~\ref{fig:raw}(A), and Proportional Integrating Measure (PIM) mode. In short, the ZCM mode measures frequency of movements and the PIM mode a quantity proportional to energy expenditure. ZCM counts the number of times per epoch that the transducer signal crosses a positive threshold close to zero; e.g., raising a hand in a single smooth movement would give 2 ZCM counts during the acceleration phase and 0 during deceleration, totaling 2 counts for the movement. In this mode high frequency artefacts may potentially be misinterpreted as considerable motion. PIM integrates the total area under the absolute value of the signal curve in a given epoch, giving an integer number in the range [0,65535] representing the sum of momentum changes. For the movement described above it would add both the acceleration and deceleration.

Raw actigraph records were cut into days and nights by finding peaks in Pearson correlation of ZCM with a step function and subsequent human verification. Individual days were excluded from the records on visual inspection due to non-compliance of some subjects to wear the device. At the group level, all days of clean data were taken into account ($9\%$ of daily records were removed due to non-compliance of some subjects to wear the device). The total numbers of analyzed daily records were 41, 33, 32, and 42 for control, combined subtype, hyperactive-impulsive subtype, and ASD, respectively. Below, whenever we refer to days and nights, we mean 14-hour diurnal and 4-hour nocturnal (before wake-up) actigraphy record, which is to normalize lengths of the samples and avoid possible artifacts due to differing sleep lengths. The details of further processing are described below.

\subsection{Selected features}
There are in total 47 features of time series that we used for classification. Here we briefly introduce them, from least to the most complex, and refer to Appendix~\ref{app:methods} for more details.
\begin{itemize}
\item 	Sleep (1-3): waking hour, hour of falling asleep and the duration of sleep.
\item	Statistics of activity (4-11), Fig.~\ref{fig:density}: we computed $2\times2\times2=8$ parameters: average and standard deviation of activity at night and during day in ZCM and PIM modes. That these quantities can be reasonable candidate parameters for prediction can be inferred from inspecting differences between activity histograms of extreme individuals visible in Fig.~\ref{fig:density}.
\begin{figure}[htbp!]
\centering
\includegraphics[width=0.98\textwidth]{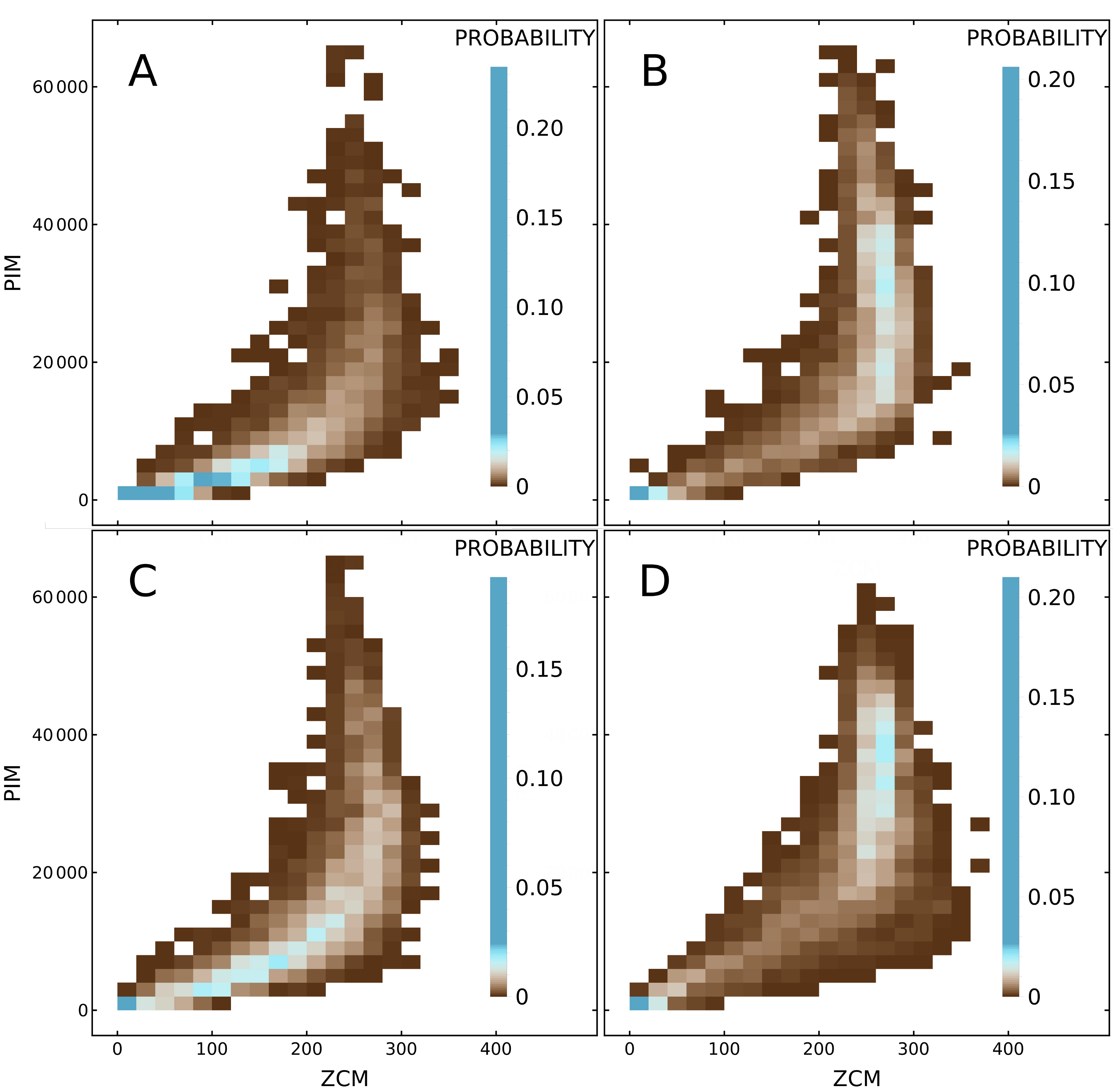}
\caption{\label{fig:density}
Density histograms of activity of the individual subjects chosen from each group. (A) control, (B) autism spectrum disorder, (C) ADHD combined type, (D) ADHD-predominantly hyperactive-impulsive type. Each colored pixel denotes how often a given subject engaged in an activity with a given Zero Crossing Mode (ZCM) and Proportional Integrating Measure (PIM) intensity. Compare with histograms for ZCM in Fig.~\ref{fig:raw}.}
\end{figure}
\item	Statistics of increments of activity (12-15): kurtosis of ZCM/PIM diurnal/nocturnal activity increments. Taking increments (i.e., differences between consecutive time points) effectively removes slow changes in a time series, which is often desirable in further analysis. The peakedness of distributions of increments, which can be quantified as kurtosis, can be observed in lower right panel in Fig.~\ref{fig:raw}.
\item	Return maps (16-19): the ZCM and PIM positions of the fixed points and the square deviation of the ZCM and PIM curves from the diagonal on the return map plot (in which we plot a series at time $t$ against the same series at $t+1$). In short, the fixed points are the states to which the system is ultimately drawn (like activity and rest in ZCM actigraphic series), and the deviations mentioned above are a proxy for the speed of returning to these states. Return maps have been used, e.g., in movement analysis in ants~\cite{Dante2017ants}.
\begin{figure}[htbp!]
\centering
\includegraphics[width=0.45\textwidth]{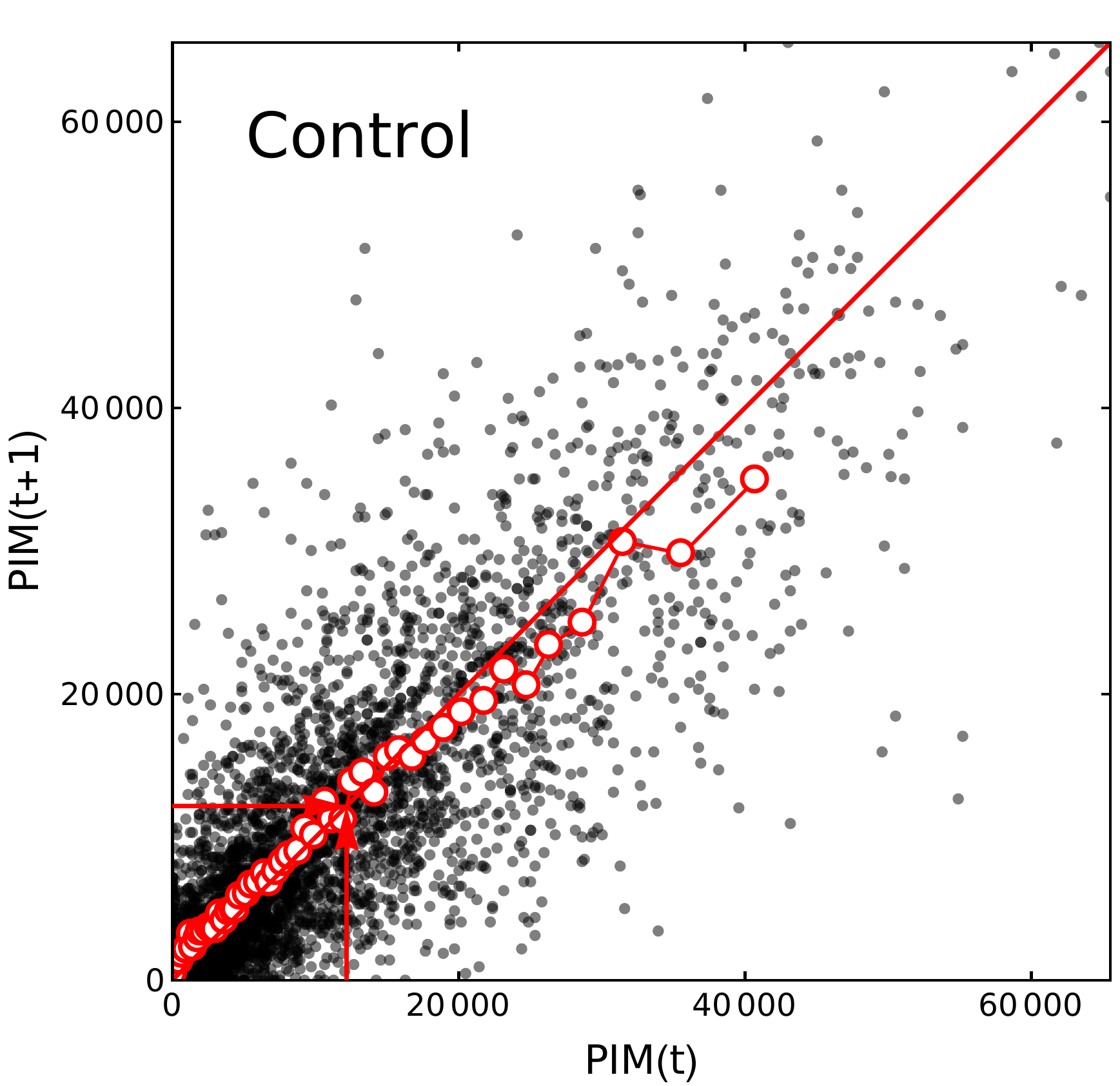}
\hfill
\includegraphics[width=0.45\textwidth]{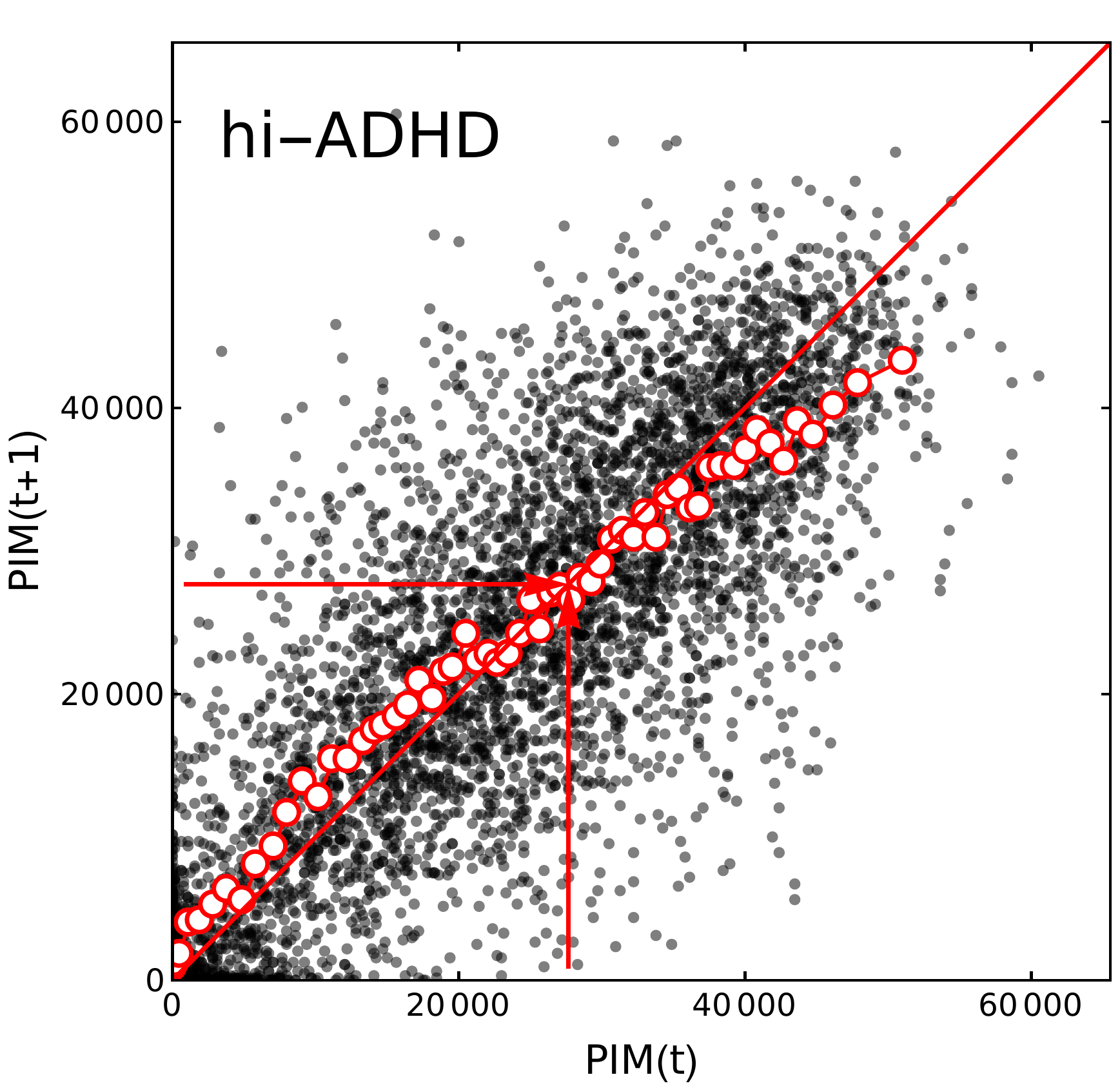}
\caption{\label{fig:return}
Return maps in PIM mode for most distinctive individuals in control and hi-ADHD groups. $x$-axis is activity at time $t$ and $y$-axis is the activity in the next time step. Red circles show averages over bins of equal number of data points. Red arrows indicate the crossing point between the map and the diagonal.}
\end{figure}
\item	Hurst exponent (20-23): subject averages of ZCM/PIM exponents over days/nights. They were computed with detrended fluctuation analysis (DFA)~\cite{Peng1994,Talkner2000}, frequently used in analysis of physiological data~\cite{Holloway2014,Rodriguez2007,Makowiec2011,Zebrowski2013,Sahin2009}. The Hurst exponent is especially known in analysis of fractal time series. It is a flagship example a measure for long-term memory of a process.
\item	Durations of activity and rest (24-31): the average duration of activity/rest, and $\alpha$ and $\gamma$ parameters for both ZCM and PIM modes. Activity and rest periods were defined as actigraphy measurements lying above or below a predefined threshold (see Fig.~\ref{fig:raw} upper panel). It has been shown that in humans and other species the durations a have complementary cumulative distributions $C(a)$ of universal shapes~\cite{Anteneodo2009,Nakamura2008}, a power-law $C(a)=a^{-\gamma}$ for rest and a stretched exponential $C(a) = \exp\left (-\alpha a^\beta \right )$ for activity periods. The parameters of these distributions can differ between conditions such as major depressive disorder, chronic sleep deprivation or schizophrenia~\cite{Nakamura2007,Nakamura2012,Ochab2014,Gudowska2016jstat}.
\item	Average shape of activity (32-35): skewness of the shapes for ZCM and PIM, and the scaling exponents $\mu$. By the shape we mean the curve ZCM(t) (or PIM(t)) taking an excursion above a predefined, low threshold. In humans and ants~\cite{Dante2015howwe,Dante2017ants} such excursions follow some universal, approximately parabolic shapes~\cite{Castellano2003,Castellano2004}, see Fig.~\ref{fig:avalanche}. The scaling exponent $\mu$ describes the scaling between durations $T$ of the excursions and their sizes $S$. The disorders under study could hypothetically manifest themselves as deviations from such universality of these quantities.
\begin{figure}[htbp!]
\centering
\includegraphics[width=0.49\textwidth]{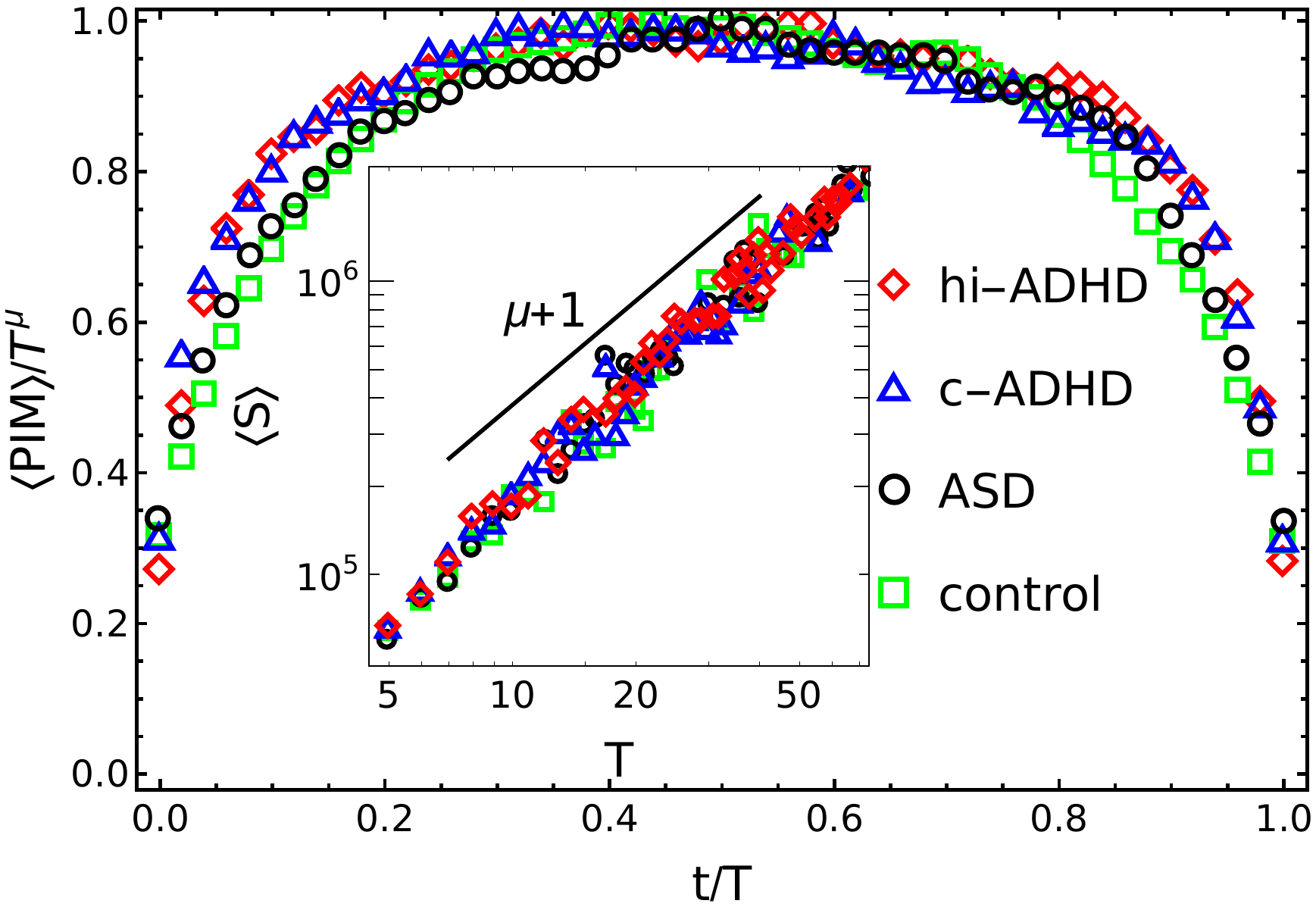}
\caption{\label{fig:avalanche} Average shape of Proportional Integrating Measure (PIM) activity (feature 8) of the most distinctive individuals from each group. Inset: universal scaling between durations $T$ of excursions and their average sizes $S$, with the scaling exponent $\mu$.}
\end{figure}
\item	Non-linear forecasting features (36-39): the skewed difference statistic~\cite{Theiler1992} and the non-linear prediction error~\cite{Farmer1988,Longtin1997} for ZCM and PIM. These measures are known in non-linear forecasting. Their intuitive meaning is as follows: the first measures the size of asymmetry between rises and falls of a numerical sequence; the second measures how well a model predicts activity at time $t+T$, given the activity at time $t$.
\item	Lempel-Ziv complexity~\cite{lempel1976complexity} (40-47): for diurnal and nocturnal ZCM and PIM activity and their increments.  This information theoretical quantity measures irregularity or randomness of a sequence, and is reduced by its repetitiveness. We used the simplest version of the algorithm for binary sequences, which involved binarization of the actigraphic records (for diurnal ZCM and PIM, substituting activity below mean with zeros and above with ones; for nocturnal records, all non-zero activity was turned into ones; the series of increments were binarized into positive and non-negative).
\end{itemize}

\subsection{Optimizing feature subset}
In this study, since both the size of the data set and the feature set are small, we performed a brute-force search for subset of features optimizing classification performance: first, we computed performance of all (1,081) feature pairs and ranked them according to their median accuracy (in a 4-class task); we chose twelve highest scoring but least correlated features; ultimately, an exhaustive search was performed over all (3,797) combinations of up to eight out of the twelve best features.

\subsection{Classifier}
Since we resorted to exhaustive search, we utilized naïve Bayes classifier due to its speed. After finding the final near optimal feature sets, we checked whether other classifiers (all implemented by default in Mathematica 11.3: nearest neighbor, logistic regression, support vector machines and random forest) improve on naïve Bayes’s performance. Ultimately, the best results that we report on, see Table~\ref{tab:best_feat}, include logistic regression, naïve Bayes, nearest neighbor.

\subsection{Cross-validation}
The sensitivity and specificity results are averages (and standard deviations) based on 100 cross-validation runs. In the 4-class (control, ASD, c-ADHD, hi-ADHD) and 3-class task (control, c-ADHD, hi-ADHD) task the scores were calculated as weighted averages over non-control classes.
In the 4-class task and 3-class task, for a given feature set each run consisted of randomly assigning four subjects from each group to the training set, and the three-four remaining ones to the test set. In the binary classification task (control versus both ADHD subtypes), in each run four control subjects, two c-ADHD and two hi-ADHD were sampled to both the training and test set, in order to retain balance between sample sizes. We classified subjects only on the level of a person (based on all available days) and not on the level of days per person, because some of the features need a larger statistic to be feasible for estimation.

\section{Results}
In the previous section we presented in total 47 features that hypothetically can give insight into the nature of hyperactivity as measured by means of actigraphy. Below we show how these features were used to train a classifier that, given a new subject, can label him or her as belonging to one of the four groups: control, ASD, c-ADHD, and hi-ADHD.

\subsection{Participants}
This study involved n=29 male participants, aged $9.89\pm0.92$ years, in four groups (7 with c-ADHD, 7 with hi-ADHD, 7 with autism spectrum disorder, ASD, and 8 in the control). There were 48 potentially eligible participants; 18 participants were excluded due to suspected co-occurring developmental disorders, and one participant was excluded due to below normal intelligence assessment. The actigraphy was recorder between in Nov-Dec 2015 (ASD), Feb and June 2016 (ADHD groups), Feb-March 2016 and one person in July 2016 (control); the time of year has marginal interaction with the classification results. The subjects in ADHD groups had been under psychological counselling for $3.4\pm2.4$ years at the time of actigraphy recording. There were no clinical interventions between or during enrollment and data collection; no medication use was reported.

\subsection{Feature correlations}
\label{sec:feat_corr}
\begin{figure}[htbp!]
\centering
\includegraphics[height=0.46\textwidth]{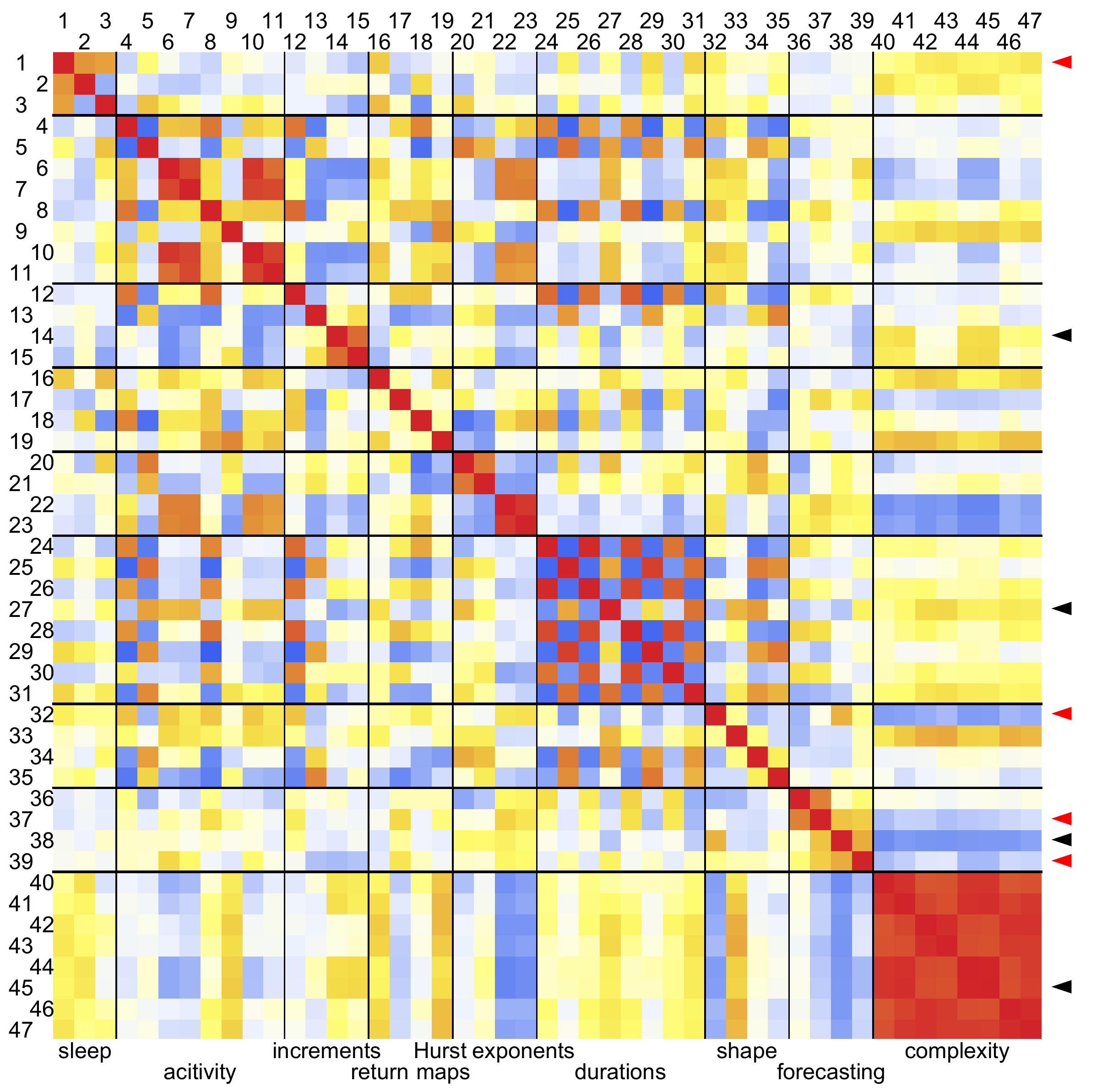}
\hfill
\includegraphics[height=0.46\textwidth]{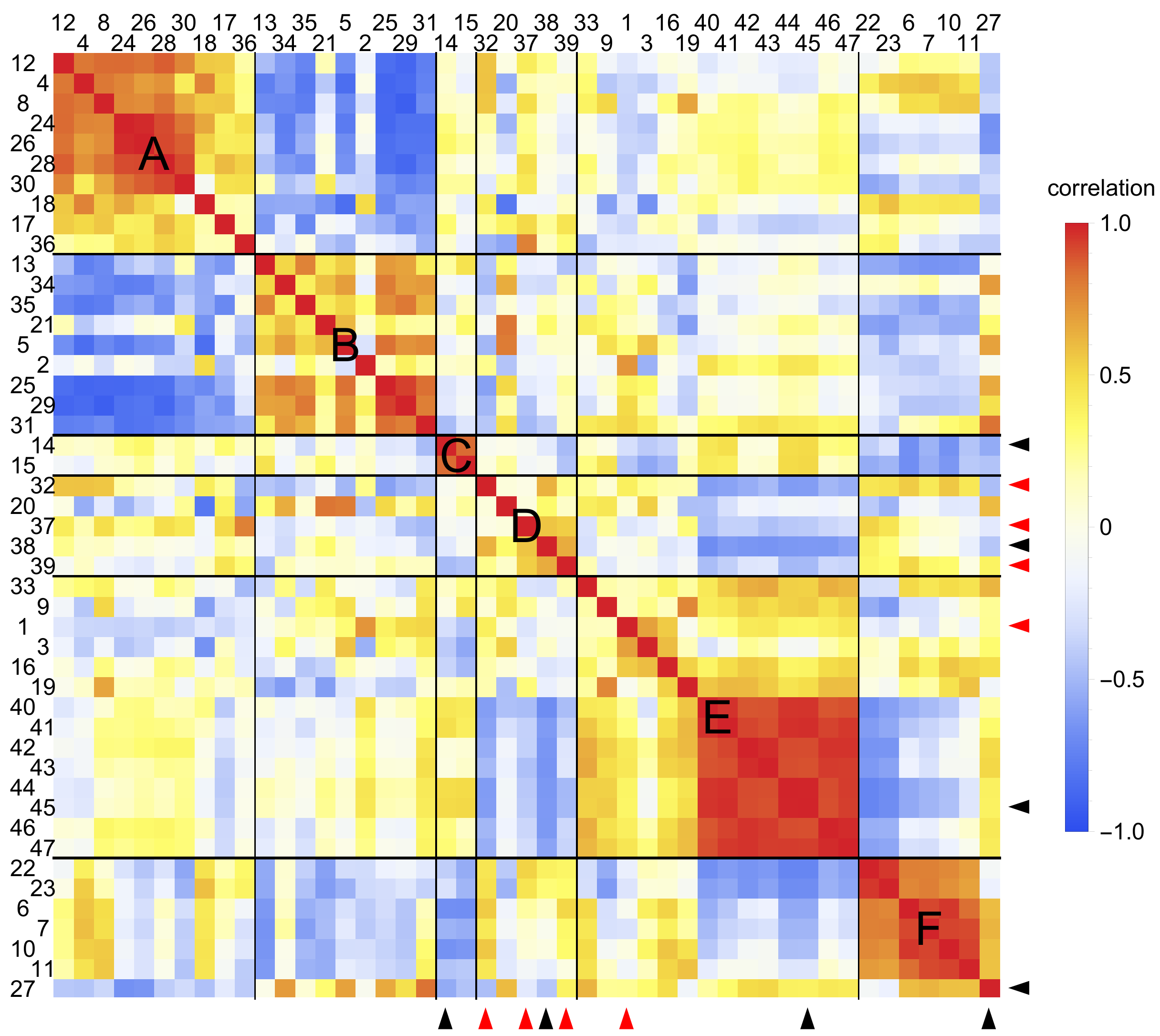}
\caption{\label{fig:corr} Left: correlation matrix of all the 47 features. Right: correlation matrix rearranged into six clusters of highly correlated features. The feature numbers, as described in text, are provided for reference. The triangles on the sides point to the eight selected features; red triangles indicate the four performing best.
}
\end{figure}

Understandably, with only 29 subjects this big number of features can lead to undesirable effects, like overfitting the classifier. However, since a lot of the features are by design strongly interrelated, we can reduce their number and obtain a better understanding of what they indicate. We used a weighted-average agglomerative clustering algorithm searching for six feature clusters, which we list and discuss below (feature numbers correspond to the order of presentation in Section “Methods: Selected features”; for visual representation of feature correlation matrix see Fig.~\ref{fig:corr}(left) below, and Fig.~\ref{fig:corr}(right) for the clustered correlation matrix):
\begin{enumerate}[label=\Alph*]
\item\label{item:1} (4, 8) mean daytime activity ZCM and PIM, (12) daytime ZCM kurtosis of increments, (17-18) return map crossing PIM and return map deviation ZCM, (24, 26, 28, 30) mean active and resting durations of ZCM and PIM, (36) skewed difference in ZCM;
\item\label{item:2} 	(2) hour of falling asleep, (13) daytime PIM kurtosis of increments, (21) daytime DFA exponent PIM, (25,29) active 
durations exponent ZCM and PIM, (31) resting durations exponent PIM, (34-35) $\mu$ exponent ZCM and PIM;
\item\label{item:3} 	(14-15) nocturnal kurtosis of ZCM and PIM increments;
\item\label{item:4} 	(20) daytime DFA exponent ZCM, (32) shape skewness ZCM,  (37-39) skewed difference in PIM and prediction error in ZCM and PIM;
\item\label{item:5} 	(1, 3) waking hour and sleep time, (8) std. dev. of daytime activity PIM, (16) return map crossing ZCM, (19) return map deviation PIM, (33) shape skewness PIM, (40-47) all Lempel-Ziv complexities;
\item\label{item:6} 	(6, 10) mean activity at night ZCM and PIM, (7, 11) std. dev. of activity at night ZCM and PIM, (22-23) night DFA exponent ZCM and PIM, (27) resting durations exponent ZCM.
\end{enumerate}

The groups A and B are very strongly anti-correlated, which for classifying purposes means they both bear similar information content. Similarly, group C is anti-correlated to group F. All the Lempel-Ziv complexities in group E are very strongly correlated with each other but at the same time they are anti-correlated with groups D and F. There is a noteworthy likeness, expected in theory, between DFA exponents, exponents of distributions of rest and activity durations, and the scaling exponents $\mu$ (cluster B). 
Notably, the highly (anti-)correlated clusters A and B do not contain features informative for classification purposes. The best features, cf. Table~\ref{tab:best_feat}, especially in cluster D, are rather weakly correlated among themselves but also to all other features.

\subsection{Best features}
The eight best performing feature subsets together with their classification algorithms, selected as described in Section “Methods: Selected features”, are listed in Table~\ref{tab:best_feat}. Most importantly, the best feature, shape skewness in ZCM mode on its own leads to a significantly non-random result with accuracy: $41.38 \pm1.1\%$ (against $25\%$ chance baseline). Combinations of remaining features incrementally enhance performance as shown in Table~\ref{tab:best_feat}. 

\begin{table}[htbp!]
\begin{tabular}{ c c | c c c c c c c c  }
\toprule
\multicolumn{2}{c|}{Classifier}&	NB &	LR &	LR &	NN &	LR &	LR &	LR &	LR \\ 
\multicolumn{2}{c|}{Sensitivity [$\%$]} &	$43.6\pm1.3$ &	$50.7\pm1.3$ &	$51.4\pm1.4$ &	$61.8\pm1.4$ &	$53.0\pm1.2$ &	$49.6\pm1.4$ &	$53.4\pm1.2$ &	$53.3\pm1.5$ \\ 
\multicolumn{2}{c|}{Specificity [$\%$]} &	$79.67\pm0.70$ &	$80.03\pm0.59$ &	$82.43\pm0.64$ &	$79.30\pm0.43$ &	$82.73\pm0.50$ &	$80.63\pm0.62$ &	$81.30\pm0.59$ &	$82.07\pm0.50$ \\ 
\hline
No & Group	 &	 &	 &	 &	 &	 & &	 & \\ 
\hline
32 &	D  &	+ &	+ &	+ &	+ &	+ &	+ &	+ &	+ \\
1 &	E   &	 &	+ &	 &	+ &	+ &	+ &	+ &	+ \\
14 &	C  &	 &	 &	+ &	 &	+ &	+ &	+ &	+ \\
39 &	D  &	 &	 &	+ &	+ &	+ &	 &	+ &	+ \\
37 &	D  &	 &	 &	 &	+ &	+ &	 &	+ &	+ \\
38 &	D  &	 &	 &	 &	 &	 &	+ &	+ &	+ \\
27 &	F  &	 &	 &	 &	 &	 &	+ &	+ &	+ \\
45 &	E  &	 &	 &	 &	 &	 &	+ &	 &	+ \\ 
\bottomrule
\end{tabular} 
\caption{\label{tab:best_feat} The eight best scoring classifiers; highest scores are in bold, and the frame indicates the selected best classifier. The values of sensitivity and specificity are averages $\pm$ standard deviation of the mean. The '+' signs in columns indicate 1-8 features included in a classifier. Classifier names: LR – logistic regression, NN – nearest neighbors, NB – naïve Bayes. Feature names: 32 – Zero Crossing Mode (ZCM) shape skewness, 1 – waking hour, 14 – nocturnal kurtosis of ZCM increments, 39 – Proportional Integrating Measure (PIM) prediction error, 37 – PIM skewed difference, 38 – ZCM prediction error, 27 – ZCM exponent of resting times $\gamma$, 45 – Lempel-Ziv complexity of PIM increments. Groups refer to correlation clusters, Sec.~\ref{sec:feat_corr}.}
\end{table}

The best nearest neighbors 4-feature classifier reached $46.5\pm1.1\%$ accuracy (against $25\%$ chance baseline), $61.8\pm1.4\%$ sensitivity and $79.30 \pm0.43\%$ specificity in a 4-class task (control, ASD, hi-ADHD, c-ADHD). The same classifier obtained $62.6\pm1.9\%$ sensitivity and $70.1\pm9.8\%$ specificity in a 3-class task (control, hi-ADHD, c-ADHD) and $69.4\pm1.6\%$ accuracy (against $50\%$ chance baseline) $78.0\pm2.2\%$ sensitivity and $60.8\pm2.6\%$ specificity in a binary task; please note that this classifier had been optimized for four-class problem. Tests on the final feature sets indicate that logistic regression systematically obtains similar scores. Due to large deviations (the provided uncertainties are standard deviations of the mean of 100 cross-validation tests), however, it is not possible to provide the one and only significantly better classifier or feature set. It is worth noting that out of the features belonging to the correlated groups A and B none were selected.

\begin{figure}[htbp!]
\centering
\includegraphics[width=0.98\textwidth]{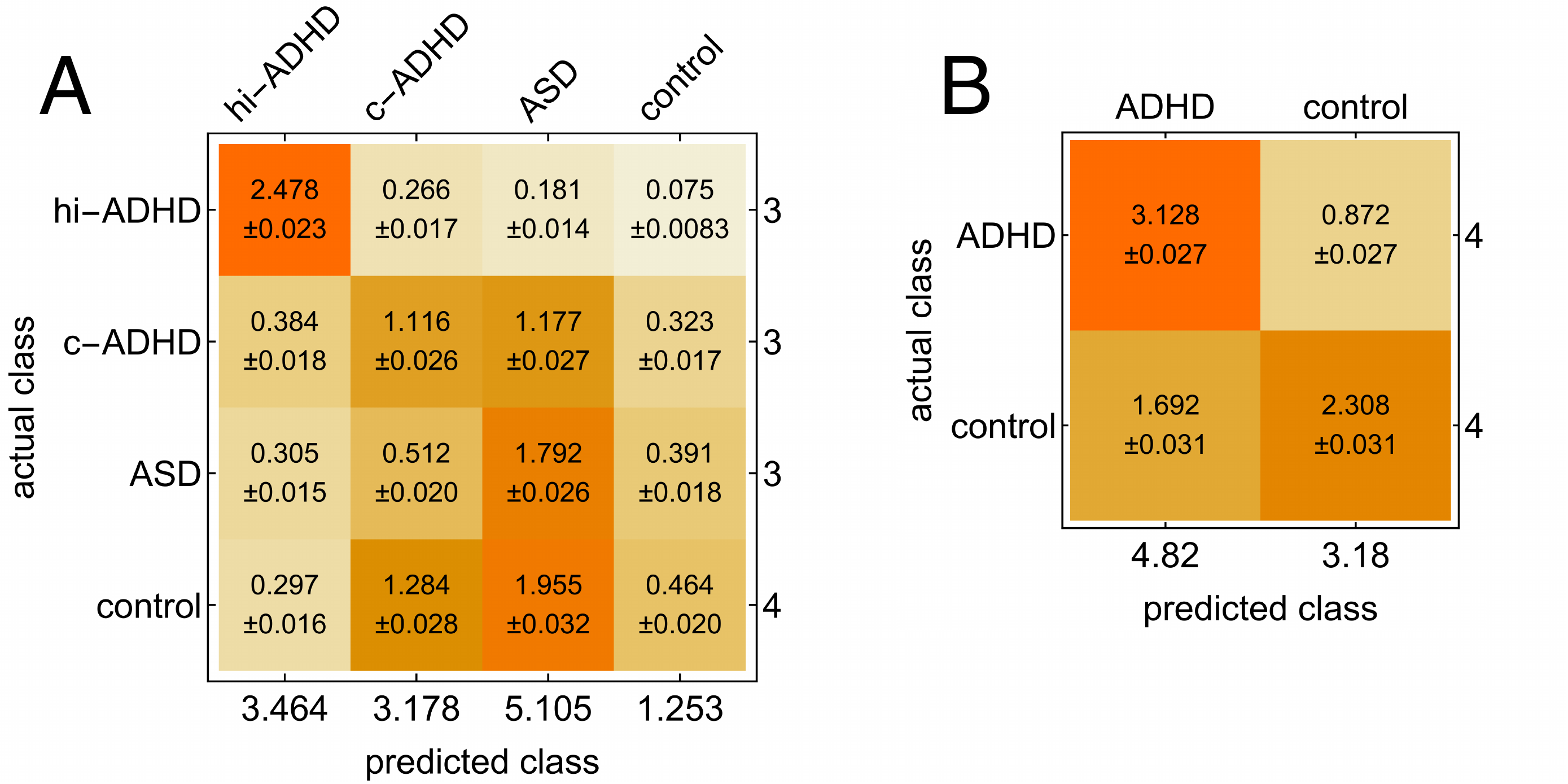}
\caption{\label{fig:confusion} (A) Confusion matrix in a 4-class task of the best 4-feature classifier; (B) confusion matrix in a binary task (control vs ADHD) with the same classifier. Numbers are means and standard deviations of the mean in a 1,000 cross-validation setup. The diagonal entries are the correctly classified cases, the off-diagonal are misclassified cases. Numbers on the axes are row and column sums.}
\end{figure}

In Fig.~\ref{fig:confusion} we supplement the results with the whole confusion matrix for the 4-feature nearest neighbors classifier using on both the 4-class and binary problem obtained after 1,000 cross-validation runs. The predictions for hi-ADHD are very good and are promising for ASD. Strikingly, the controls are often misclassified as either ASD or c-ADHD condition, and the c-ADHD group is in turn often misclassified as ASD. This serves as a cautionary example that some features might not be specific to one disorder only. In the binary case, the classifier is highly sensitive to the ADHD condition.

\section{Discussion and conclusions}
The results presented in this paper indicate that advanced analysis of actigraphy measurements can enhance diagnostics of hyperactivity disorders in children. In a realistic setting patients may differ in type and intensity of disorder symptoms (like ADHD subtypes), as well as they can actually suffer from another disorder (here, the ASD group was used as a convenient way of further confusing the classifiers). For these reasons we believe that a classifier successfully tested on multiple classes is reliable and it might be general enough to distinguish in a controlled setting also other disorders.

Merely one quantity is enough to produce results better than chance: the skewness of the averaged shape of activity of a person. The results show that also other selected quantities are informative, e.g., non-linear forecasting measures (prediction error and skewed difference), the peakedness of the histogram of differenced nocturnal activity, or simply a person’s waking hour. These might offer clues as to the nature of symptoms in hyperactivity disorders. They can also already improve diagnosis of these disorders, especially with the use of wearable telemetric devices.

Further experimental studies should explore how to observe differences in symptom intensity in children with ADHD combined subtype and hyperactivity-impulsiveness subtype as they relate to the changes in parameters and topography in both hemispheres of the event-related potential connected to the process of categorizing new and known stimuli. This proposed context of differentiation between ADHD subtypes would not only be more precise, but would also provide new interdisciplinary perspectives for therapeutic interventions for this disorder.

Actigraphy results, supported by appropriate mathematical interpretations, could be useful for detecting subclinical deficits in the subtypes. Analysis of the obtained results showed that such a method can effectively differentiate between children with specific ADHD subtypes and those without ADHD, and that it allows for the detection of subtle disorders of motor activity, as well as, indirectly, cognitive processing abilities. Therefore, it seems legitimate to consider actigraphy in addition to the analysis of event-related potentials when diagnosing ADHD, in order to make an unambiguous diagnosis, and to avoid running into serious difficulties when preparing to provide appropriate and effective therapeutic intervention strategies.

\begin{acknowledgements}
JKO thanks Valeria Pattacini from the Office of International Relations of Universidad de San Mart\'{i}n, UNSAM, (Argentina) for facilitating his visit and the UNSAM’s hospitality. 
We would like to thank Anna Bereś at the Institute of Applied Psychology, Jagiellonian University, for discussions on study design at the initial stages of the project, and Maria Marczyk from “Effatha” Centre for Autistic People, for the acquisition of data from ASD-diagnosed children. 
Work conducted under the auspice of the Jagiellonian University-UNSAM Cooperation Agreement. JKO was supported by the Grant DEC-2015/17/D/ST2/03492 of the National Science Centre (Poland). DRC was supported in part by CONICET (Argentina) and Escuela de Ciencia y Tecnolog\'{i}a, UNSAM.
\end{acknowledgements}

\appendix

\section{Methods supplement}
\label{app:methods}

%
%




\subsection*{Sleep}

Three straightforward quantities connected with sleep were taken into account: waking hour, hour of falling asleep and the duration of sleep. 
Obviously, they are dependent, but it is not a priori clear which of them can be most informative.
The differences between the groups indicated in Tab.~\ref{tab:sleep} show that they can be of moderate use at the prediction stage.

\begin{table}[htbp!]
\begin{tabular}{r|c c c|c c c|c c c}
\toprule
 & \multicolumn{3}{c|}{Waking} & \multicolumn{3}{c|}{Falling asleep} & \multicolumn{3}{c}{Sleep duration}\\
 &  ASD & c-ADHD & hi-ADHD  &  ASD & c-ADHD & hi-ADHD  &  ASD & c-ADHD & hi-ADHD \\ 
\hline 
control & 0.63 & 0.20 & 0.078 & 0.23 & 0.98 & 0.060 & 0.11 & 0.14 & 0.39 \\
ASD &  & 0.19 & 0.33 &   & 0.19 & \textbf{0.0061} &   & \textbf{0.026} & 0.054 \\
c-ADHD &  &   & \textbf{0.018} &  &   & \textbf{0.039} &  &   & 0.53 \\
\bottomrule
\end{tabular} 
\caption{\label{tab:sleep} P-values ($<0.05$ in bold) of two-sided Student's t-tests for differences between the experimental groups in waking time, time of falling asleep and the duration of sleep.}
\end{table}

\subsection*{Statistics of activity}

The starting point of the data analysis was a visual inspection of activity histograms of the extreme individuals visible in Fig.~\ref{fig:density}.
The activity included five days trimmed to 4 hours of sleep (before wake-up) followed by 14 hours of daytime (the trimming is performed to avoid possible effects due to differing sleep lengths, which are treated separately).
It is easy to notice that ADHD subject (and similarly the ASD subject) has a relatively higher volume of activity around $ZCM=260$ and $PIM=20 000-45 000$, while the main control subject's activity lies in the region $ZCM<200$ and $PIM<10 000$, and the c-ADHD subject is intermediate between the two. The high probability in the yellow pixel around $0$ is a result of the low nocturnal activity.

The above observations allow to assume that mean and standard deviation of activity (that is roughly the location and spread of activity in Fig.~\ref{fig:density}) 
are reasonable candidate parameters for prediction.
Thus we computed $2\times 2 \times 2 = 8$ parameters: average and standard deviation for night and day in ZCM and PIM mode separately.
For each subject the average and standard deviation were calculated from concatenation of all the valid 4-hour (or 14-hour) periods.

Additionally, when examining the set of all the subjects with each day and night treated separately, the mean presented strong anticorrelations with variance in ZCM day, and strong correlation in ZCM and PIM nights, see Table~\ref{tab:corr}. This behavior can be understood in terms of the universal shape of the activity histograms (e.g., at night whenever non-zero activity occurs, the variance naturally rises).

\begin{table}[htbp!]
\begin{tabular}{r|c|c|c|c}
\toprule
 & control & ASD & c-ADHD & hi-ADHD \\ 
\hline 
ZCM day < & -0.61 & -0.70 & -0.61 & -0.74 \\ 
ZCM night > & 0.95 & 0.65 & 0.76 & 0.92 \\ 
PIM night > & 0.94 & 0.75 & 0.85 & 0.93 \\ 
\bottomrule
\end{tabular} 
\caption{\label{tab:corr} Correlations between mean and variance of activity; p-value $<0.05$, Spearman one-sided test for correlation coefficient being lower/greater than the given value. For PIM day the correlations are insignificant or weak.}
\end{table}

\subsection*{Statistics of increments of activity}

Removing slow changes in a time series can be helpful in further analysis.
It can be achieved simply by differencing the raw time series, resulting in a series of activity increments, as shown in Fig.~\ref{fig:raw}.
It appears that not only the histograms of raw activity differ between the groups, as discussed earlier, but also the histograms of increments.
As shown in Fig.~\ref{fig:raw}
(lower right panel), the extreme subjects from different groups exhibited varying peakedness of such histograms.
This can be simply measured by kurtosis of the increment distributions.
Consequently, we included another four features for classification purposes: kurtosis of PIM/ZCM diurnal/nocturnal differenced activity.


\subsection*{Return map}

Based already on Fig.~\ref{fig:density}
it is visible that one can define two states, especially for hyperactive subjects: resting (around zero ZCM and PIM values) and active (spread around $ZCM=260$ and $PIM= 30 000$).
These points can be found by means of so called return (or Poincar\'{e}) maps that have been used, e.g., in movement analysis in ants \cite{Dante2017ants}.

Given a series at time $t$, we plot it against the same series one step forward at $t+1$, see Fig.~\ref{fig:return}.
If the obtained line were a diagonal, it would mean that starting from any activity at time $t$ the subject would remain with the same value of activity in the next time step.
If the obtained line crosses the diagonal, as it does in Fig.~\ref{fig:return} around $PIM=30000$ for ADHD, it is a fixed point of the activity. In this case, it means that the activity in some range below and above ultimately makes its way towards that point, and the speed of that travel depends on the crossing angle.
Mark that there is another fixed point at null activity.

From these considerations another four features were computed:
the ZCM and PIM positions of the crossing (of an interpolated curve with the diagonal), and the square deviation of the curves from the diagonal (as a more robust proxy for the estimate of the crossing angle).

\subsection*{Hurst exponent}

We further consider properties known from analysis of fractal time series.
A flagship example is the Hurst exponent, measuring long-term memory of a process.
It can be estimated by several interrelated methods like Fano factor and Allan variance (see, e.g.,~\cite{Thurner1997,Anteneodo2010,Gudowska2016jstat}), wavelet analysis and other ones.
We decided to turn to the detrended fluctuation analysis (DFA)~\cite{Peng1994,Talkner2000}, frequently used in analysis of physiological data~\cite{Holloway2014,Rodriguez2007,Makowiec2011,Zebrowski2013,Sahin2009}.

The scaling exponents were obtained separately for each 14-hour day and 4-hour night. The basic DFA variant was computed with only linear trends removed.
For a given subject, the features used were: average of ZCM/PIM exponents over days/nights.


\subsection*{Durations of activity and rest}

Another set of properties of interest -- very much related to the concept above -- concerns inter-event periods~\cite{Bickel1999,Anteneodo2010}.
In particular, if an event is defined as the moment of activity crossing a given threshold, the inter-event periods become the durations of high activity and rest (above and below the threshold, respectively).
It has been shown that in humans and other species the durations $a$ have complementary cumulative distributions $C(a)$ of universal shapes~\cite{Anteneodo2009,Nakamura2008}, a power-law
$C(a) = a^{-\gamma}$
for rest and a stretched exponential
$C(a) = \exp\left (-\alpha a^\beta \right )$
for activity periods; the precise form of the distribution can be further improved~\cite{Breakspear2017}.
Moreover, the parameters of these distributions can differ between conditions such as major depressive disorder, chronic sleep deprivation or schizophrenia~\cite{Nakamura2007,Nakamura2012,Ochab2014,Gudowska2016jstat}.

There are eight features that were extracted from inter-event period lengths:
the average duration of activity/rest, and $\alpha$ and $\gamma$ parameters for both ZCM and PIM modes.
The first quantity, the average period duration, has been used~\cite{Anteneodo2009,Nakamura2008} as a rescaling factor that allowed distributions to conform to a universal shape; thus in itself it might be indicative of individual characteristics of the time series.
Here we computed it for each day separately and averaged over all days for a given subject.
The parameters $\alpha$ and $\gamma$ were fitted using log-log or log-linear data for power-law and stretched exponential, respectively, in order to account for the tails in the distributions.
For a better stability of the stretched exponential fit the other parameter was set to a constant, rough estimate $\beta=0.5$.
For a given subject, the fit was performed on joint data from all available days.

\subsection*{Average shape of activity}
\label{sec:aval}

Apart from all the statistics described above, it is reasonable to assume that simply the shape the time course of activity can differ between individuals or conditions.
By the shape we mean the curve $ZCM(t)$ (or $PIM(t)$) taking an excursion above a predefined threshold. 
It has been shown for humans and ants~\cite{Dante2015howwe,Dante2017ants} that such excursions follow some universal, approximately parabolic shapes~\cite{Castellano2003,Castellano2004}, see Fig.~\ref{fig:avalanche}.
To obtain this result one has to first determine the scaling between durations $T$ of the excursions and their sizes $S$.
Upon visual inspection it became apparent that despite similar form the shapes can be skewed towards beginning or the end of the excursion.

We therefore included as additional four features: skewness of the shapes for ZCM and PIM, as well as the scaling exponents $\mu$.
The quantities were computed with $ZCM=100$ and $PIM=10000$ activity thresholds, and jointly out of all available days for a given subject.

\subsection*{Prediction error}
Furthermore, we compute two measures known in non-linear forecasting:
one is the skewed difference statistic \cite{Theiler1992}
\beq
\label{eq:Q}
Q(T) = \frac{\langle \left(x_{t+T} - x_t \right)^3 \rangle}{\langle \left(x_{t+T} - x_t \right)^2 \rangle},
\eeq
and the other is the non-linear prediction error \cite{Farmer1988,Longtin1997}
\beq
\label{eq:preErr}
E^2(T) = \frac{\langle \left(f_{t,T} - x_{t+T} \right)^2 \rangle}{\langle \left(x_{t} - \bar{x} \right)^2 \rangle},
\eeq
where $x_t$ is a given time series, $\bar{x}$ is its time average, $T$ is the number of time steps ahead we try to predict, and $f_{t,T}$ is the function we construct to predict the behavior at time $t+T$ given some knowledge at time $t$.

The intuitive meaning for the first measure is the size of asymmetry between rises and falls. In the simplest case, linearly increasing activity produces $Q(T)\sim T$, while linearly decreasing activity $Q(T)\sim -T$. If an activity is roughly constant, but it tends to rise up or fall down after some given time $T$, this would likewise be reflected in $Q$.
Since one of our hypotheses is that the shape of activity might differ between healthy and disordered people, as already indicated in Sec.~\ref{sec:aval}, it is a valid measure to investigate.

\begin{figure}[htbp!]
\centering
\includegraphics[width=0.49\textwidth]{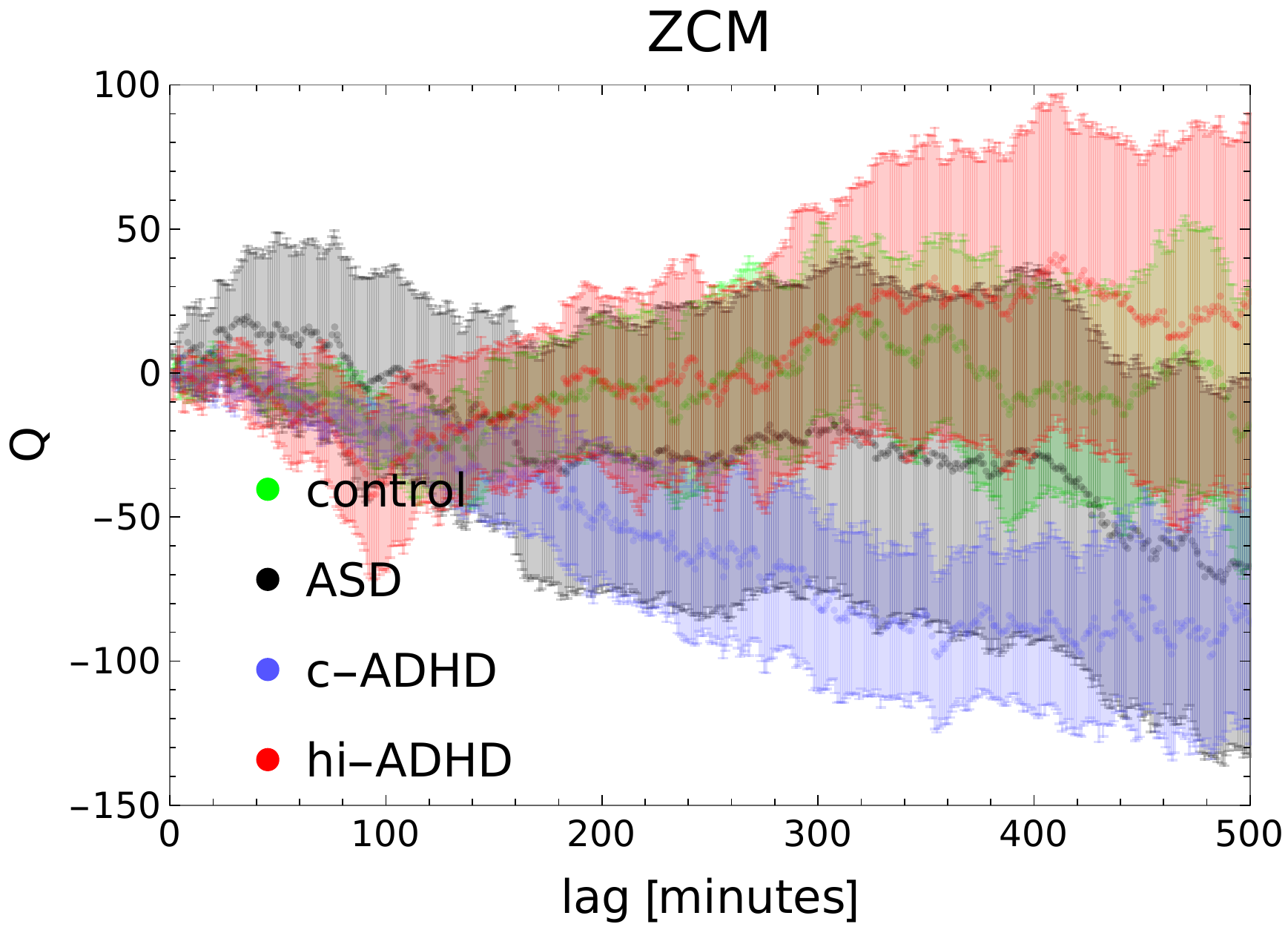}
\hfill
\includegraphics[width=0.49\textwidth]{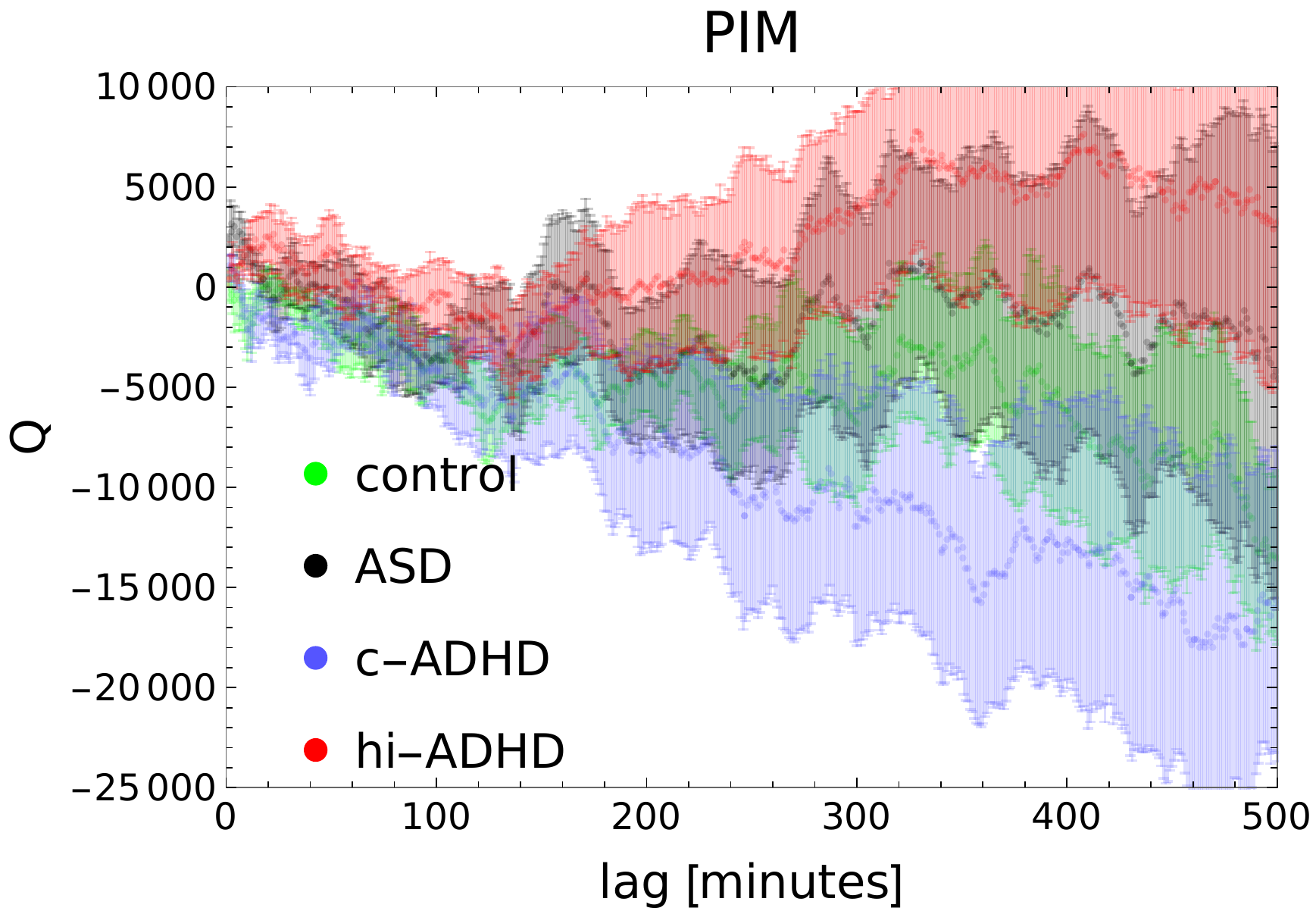}
\vfill
\includegraphics[width=0.49\textwidth]{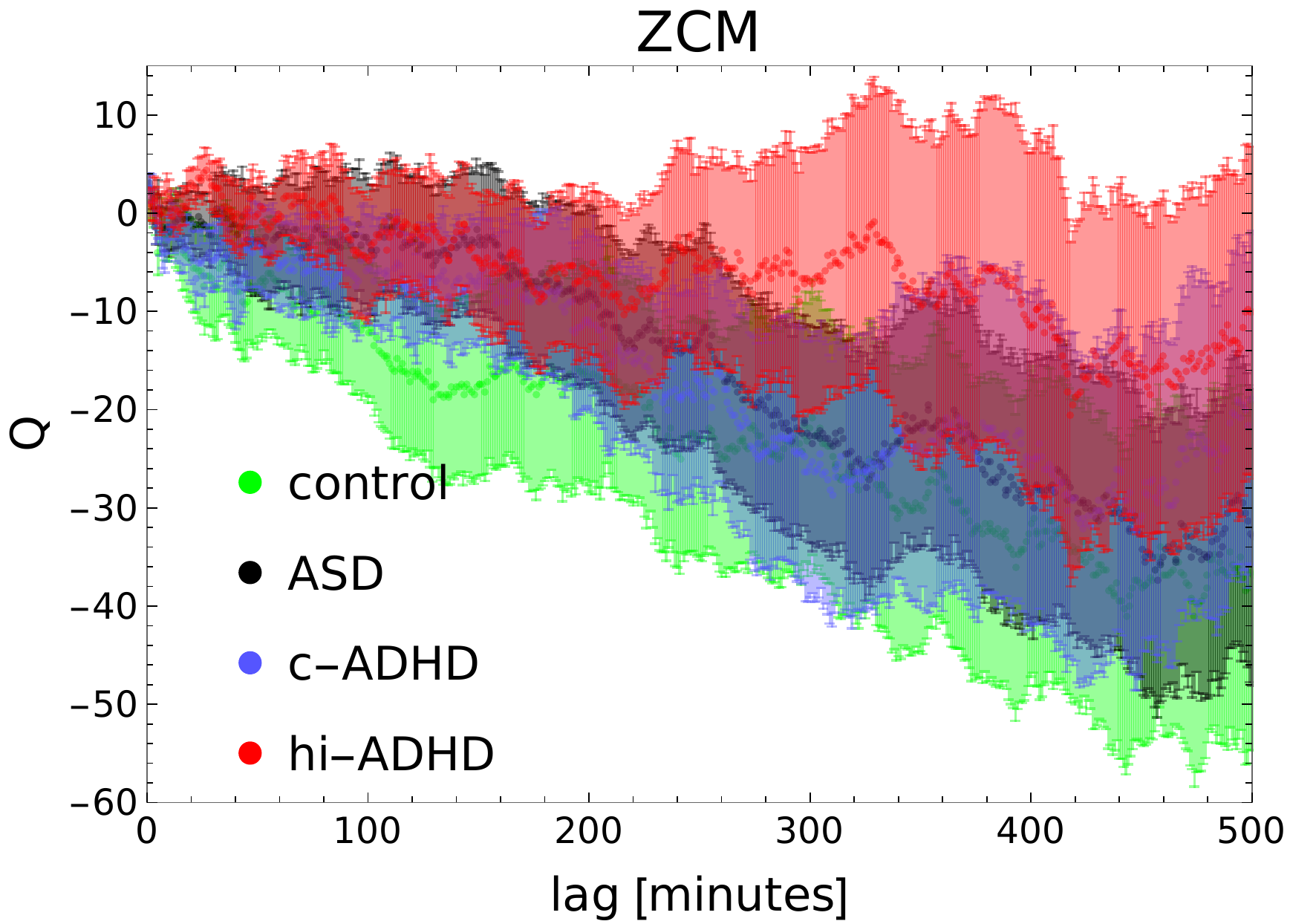}
\hfill
\includegraphics[width=0.49\textwidth]{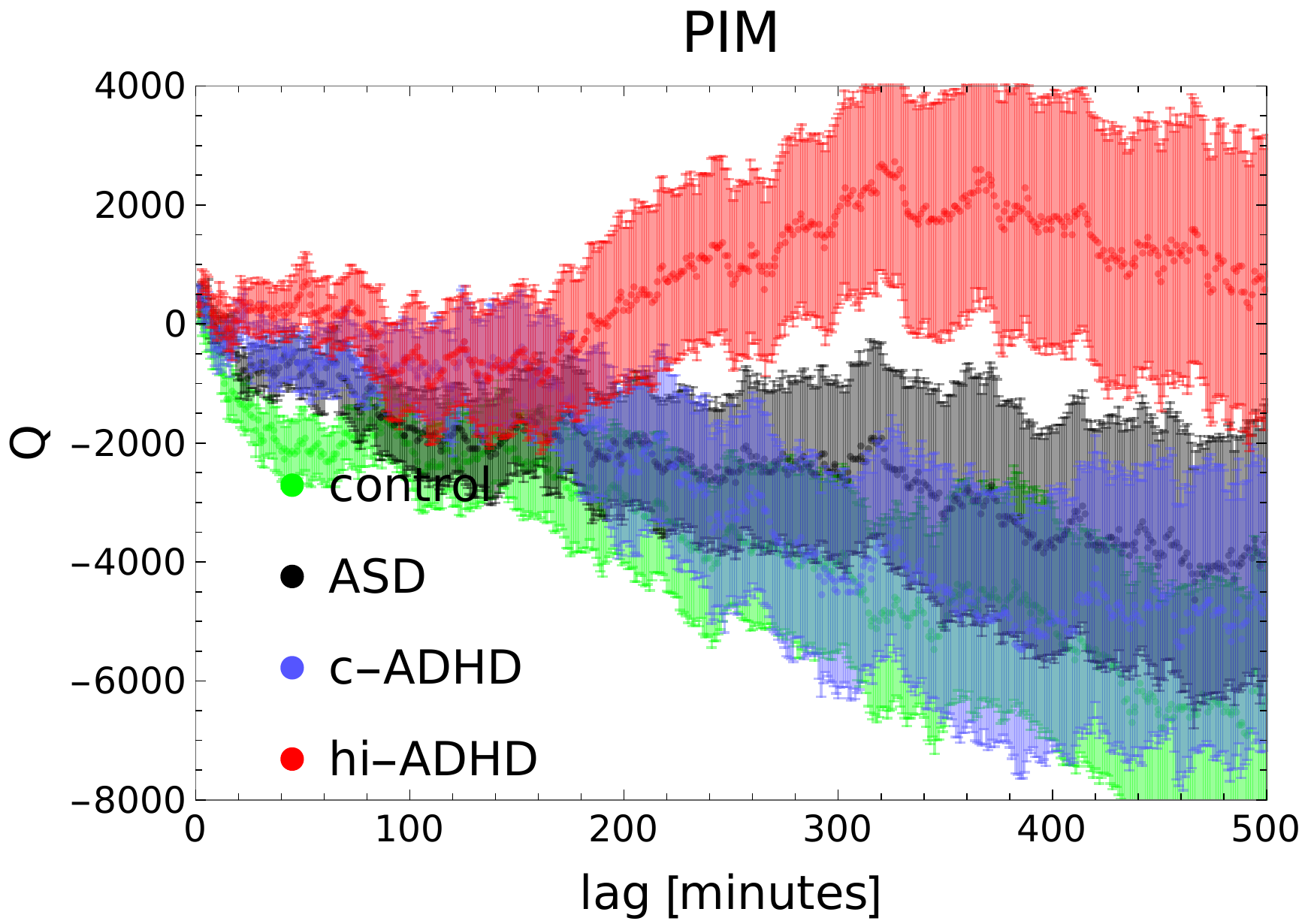}
\caption{\label{fig:Q} Skewed difference \eqref{eq:Q} for ZCM and PIM modes on (upper panels) single individuals and (lower panels) group level. From each day only  12-hour long stretch of daily activity was taken.
}
\end{figure}

The prediction error is technically much more involved, although simple conceptually: given some activity at time $t$ we want to predict it at time $t+T$, we construct a model with the prediction $f_{t,T}$ and check it against the real future activity. If the model is simply the average activity $\bar{x}_{t}$, the error will be equal to $E^2=1$; if the prediction is perfect, $E^2=0$.

Note that our goal here is actually not to optimize the predictions or to test for non-linearity, but to simply examine if for a given model the prediction error will be informative of the activity disorder.
To that purpose we use a somewhat arbitrary, but a very straightforward following model.
We divide the recordings into training and test set; we take an hour-long piece $X_t=\{x_t'\}_{t'=t-60}^{t}$ from the test set and ask for a prediction 20 minutes in the future $\{x_{t'}\}_{t'=t+1}^{t+20}$; in the training set, we find an ensemble of $k=5$ hour-long pieces which look most similar to $X_t$, which we call nearest neighbors, $\{Y_{i,t}\}_{i=1}^{k}$. The prediction is given simply by the average $f(t,T)=1/k \sum_{i=1}^{k}y_{i,t+T}$.

\begin{figure}[htbp!]
\centering
\includegraphics[width=0.49\textwidth]{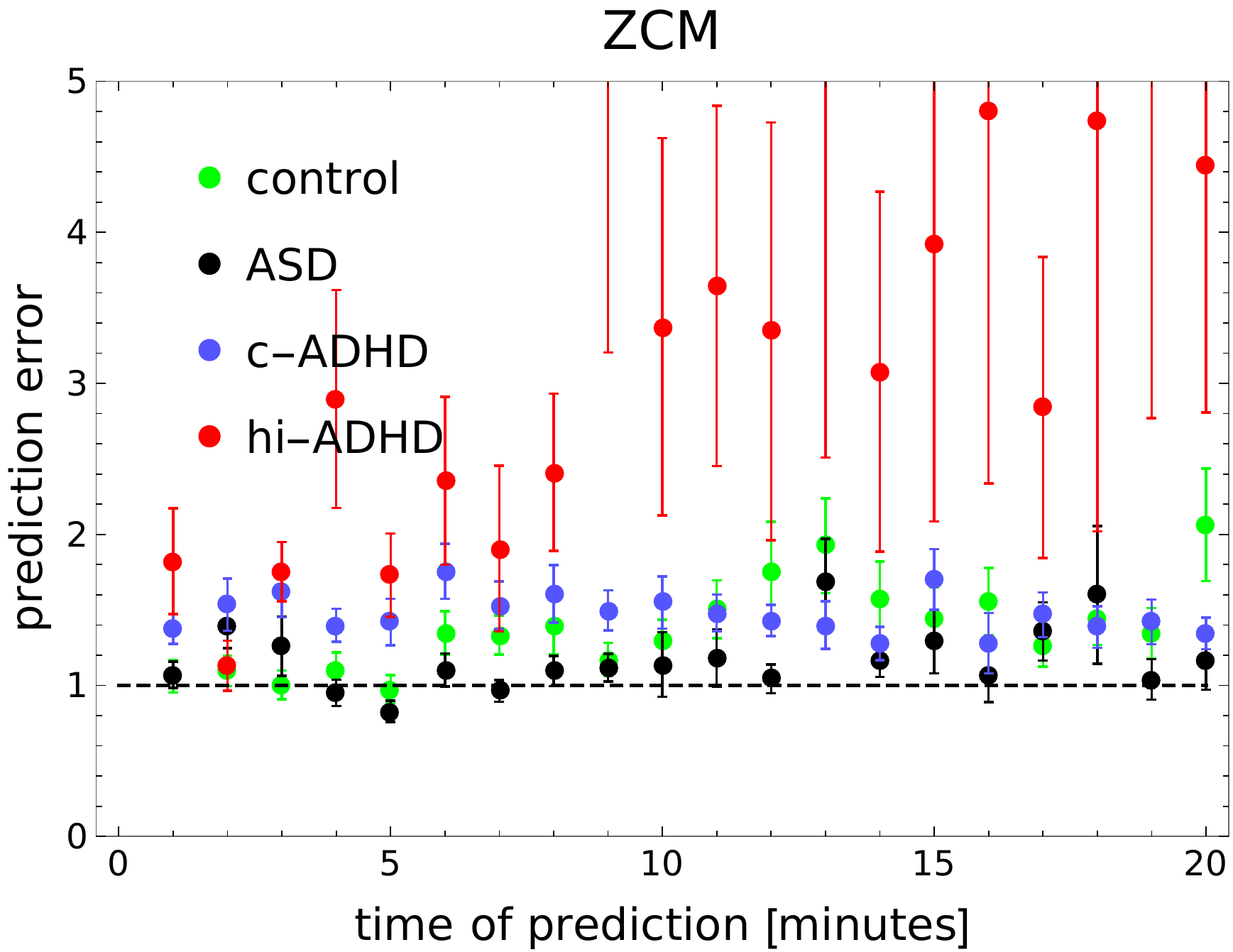}
\hfill
\includegraphics[width=0.49\textwidth]{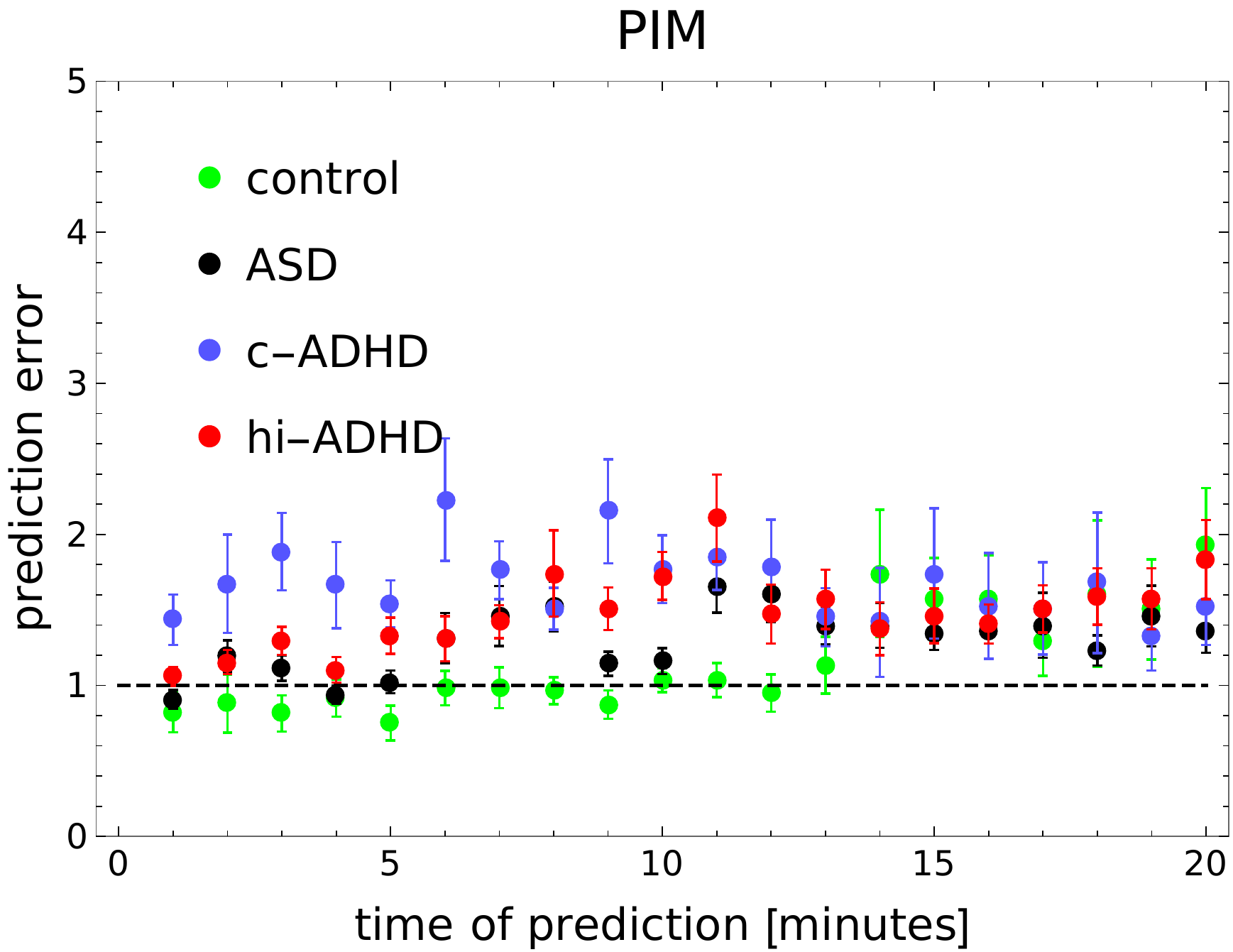}
\caption{\label{fig:E2} Prediction error \eqref{eq:preErr} for ZCM and PIM modes on most distinctive individuals. From each day only  10-hour long stretch of daily activity was taken. The prediction error values above three in the hi-ADHD subject in ZCM are not representative of the whole group. 
}
\end{figure}

Both the skewed difference and prediction error are functions of $T$.
For the purpose of classification, instead of functions we preferred single numbers, so we used average scores (over the range $T=1-500$ for $Q(T)$ and $T=1-20$ for $E^2(T)$, respectively, as presented in Fig.~\ref{fig:Q}-\ref{fig:E2}).
For a given subject in ZCM/PIM mode, the features were calculated for each day separately and subsequently were averaged.


%

%
%

\end{document}